\documentclass[aps,prc,preprint,showpacs,superscriptaddress]{revtex4}
\usepackage{epsfig}
\usepackage{amsmath}
\usepackage[hyperref]{hyperref}

\begin{document}

\title{Medium modifications of the bound 
nucleon generalized parton distributions and the quark contribution to the spin sum rule}

\author{V. Guzey}
\email{vguzey@jlab.org}
\affiliation{Theory Center, Thomas Jefferson National Accelerator Facility, 
Newport News, VA 23606, USA}

\author{A.W. Thomas}
\email{awthomas@jlab.org}
\affiliation{Theory Center, Thomas Jefferson National Accelerator Facility, 
Newport News, VA 23606, USA}
\affiliation{College of William and Mary, Williamsburg, VA 23178, USA}

\author{K. Tsushima}
\email{tsushima@jlab.org}
\affiliation{Excited Baryon Analysis Center (EBAC) and Theory Center,
Thomas Jefferson National Accelerator Facility, 
Newport News, VA 23606, USA}

\preprint{JLAB-THY-09-943}

\begin{abstract}

We estimate the nuclear medium modifications
of the quark contribution to the bound nucleon spin sum rule, $J^{q^{\ast}}$,
as well as the separate helicity, $\Delta \Sigma^{\ast}$, and the angular 
momentum, $L^{q^{\ast}}$, contributions to $J^{q^{\ast}}$. 
For the calculation of the bound nucleon generalized parton distributions (GPDs), 
we use as input the bound nucleon 
elastic form factors predicted in the quark-meson coupling model.
Our model for the bound nucleon GPDs is relevant for incoherent
deeply virtual Compton scattering (DVCS) with nuclear targets.
We find that the medium modifications increase  $J^{q^{\ast}}$
and $L^{q^{\ast}}$ and decrease $\Delta \Sigma^{\ast}$
 compared to the free nucleon case. 
The effect is large and increases with increasing nuclear density $\rho$.
For instance, at $\rho=\rho_0=0.15$ fm$^{-3}$, 
$J^{q^{\ast}}$ increases by 7\%,  $L^{q^{\ast}}$ increases by 20\%, and
$\Delta \Sigma^{\ast}$ decreases by 17\%.
These in-medium modifications of the bound nucleon spin properties 
are a general feature of relativistic mean-field quark models and
may be understood qualitatively in terms of the enhancement 
of the lower component of the quark Dirac spinor in the
nuclear medium.

\end{abstract}

\maketitle


Properties of hadrons in a nuclear medium are expected to be modified compared to those
in a vacuum. This manifests itself in the modifications of quark and gluon parton distributions of the bound nucleon measured in deep inelastic scattering (DIS) with unpolarized nuclear targets~\cite{EMC,Frankfurt:1988nt,Arneodo:1992wf,Geesaman:1995yd,Piller:1999wx}.
Even stronger medium modifications have been predicted for DIS with polarized nuclear targets~\cite{Guzey:1999rq,Bissey:2001cw,Cloet:2006bq}.
Possible medium modifications of the bound-nucleon elastic form factors were 
probed by the polarization transfer measurement in the
$^4{\rm He}(\vec{e},e^{\prime}\vec{p})^3{\rm H}$ reaction at 
the Hall A Jefferson Lab experiment~\cite{Strauch:2002wu,Malace:2008gf}.
The results of the experiment have been described by either the 
modified elastic form factors as predicted by the quark-meson coupling (QMC) model~\cite{Lu:1997mu} or by the strong 
charge-exchange final-state interaction (FSI)~\cite{Schiavilla:2004xa}.
However, such a strong FSI may not be consistent with the induced 
polarization data---see Ref.~\cite{Malace:2008gf} for details.
In addition to the modification of structure functions (parton distributions)
and elastic form factors of the bound nucleon, various static properties of hadrons 
(masses, magnetic moments, coupling constants) have been  predicted
to be modified in a nuclear medium (see e.g.,~\cite{Saito:2005rv}).

Generalized parton distributions (GPDs) interpolate between parton distributions and
elastic form factors~\cite{Ji:1998pc,Goeke:2001tz,Diehl:2003ny,Belitsky:2005qn}.
Therefore, it is natural to expect that GPDs of the bound nucleon should also be 
modified in the nuclear medium. 
An early investigation~\cite{Liuti:2005gi,Liuti:2006dx} of such modifications in $^4$He
assumed that in-medium nucleon GPDs are modified through the kinematic off-shell effects 
associated with the modification of the relation between the struck quark's transverse 
momentum and its virtuality. 
Recently, we considered 
incoherent deeply virtual Compton scattering (DVCS) on $^4$He, $\gamma^{\ast}\, ^4{\rm He} \to \gamma p X$,
and suggested a model of the bound
nucleon GPDs in $^4$He, where the GPDs are modified in proportion to the
corresponding bound nucleon elastic form factors~\cite{Guzey:2008fe}.
In the present work, we extend our approach to an arbitrary 
nucleus (any nuclear density) and study the medium modifications of the quark contribution to Ji's spin sum rule~\cite{Ji:1996ek}, $J^{q^{\ast}}$.
As in our recent work~\cite{Guzey:2008fe}, the present model of the bound nucleon GPDs
is relevant for incoherent DVCS (and other incoherent exclusive processes) with nuclear targets, $\gamma^{\ast}\, A \to \gamma N X$, 
where $A$ denotes the nucleus,  $N$ is the final detected nucleon, and
$X$ is the undetected product of the nuclear breakup. 
We find that medium modifications increase 
$J^{q^{\ast}}$ and the effect is quite noticeable.
The effect increases with increasing nuclear density $\rho$.
For instance, at $\rho=\rho_0=0.15$ fm$^{-3}$ 
($\rho_0$ is the density of the nuclear matter or, 
to a good accuracy, the density in the center 
of a nucleus), the increase is 7\%.
Separating $J^{q^{\ast}}$ into the quark helicity contribution, $\Delta \Sigma^{\ast}$,
and the quark orbital momentum contribution, $L^{q^{\ast}}$, we find that
the medium modifications decrease $\Delta \Sigma^{\ast}$ and increase $L^{q^{\ast}}$.
At $\rho=\rho_0=0.15$ fm$^{-3}$, $\Delta \Sigma^{\ast}$ decreases by 17\%
and $L^{q^{\ast}}$ increases by 20\%.

Before presenting details of our calculations,
we explain that  modifications of the bound nucleon
spin properties in the nuclear medium  
may be understood in terms of the enhancement 
of the lower component of the quark  wave function
in the nuclear medium, which is a general feature of relativistic mean-field quark models 
and which has the following consequences.
\begin{itemize}
\item[(i)]
The axial coupling constant of the nucleon is suppressed in the nuclear medium,
$g_A^{\ast} < g_A$, where the quantities with an asterisk
refer to the in-medium nucleon and the quantities without one refer to the free nucleon. 
The suppression of $g_A$  was deduced from the measurements of the nuclear Gamow-Teller beta decay~\cite{Buck:1975ae,Goodman:1985xi,Wilkinson:1973zz,Chou:1993zz}
and confirmed by theoretical calculations using the Nambu--Jona-Lasinio model~\cite{Cloet:2006bq},
the quark-meson coupling model~\cite{QMCgA,QMCneuA}, and chiral 
perturbation theory~\cite{Vaintraub:2009mm}.
The suppression of $g_A$ and of the axial vector form factor can be explained
by the Lorentz structure of the axial current and by the enhancement 
of the lower component of the quark  spinor in the nuclear medium.
In the framework of relativistic mean-field quark models---we use the results of the
quark-meson coupling model~\cite{QMCgA,QMCneuA}---the mechanism of the suppression is independent 
of the isospin structure of the corresponding matrix element.
Therefore, similarly to the suppression of the isovector axial coupling constant $g_A$,
it is also predicted that
the isoscalar quark helicity 
contribution to the bound nucleon spin, $\Delta\Sigma^{\ast}$,
is suppressed compared to that in the vacuum,
$\Delta\Sigma$ 
[for the definition of $\Delta\Sigma^{\ast}$, see Eq.~(\ref{eq:delta_q})].
Therefore,
\begin{equation}
g^{\ast}_A < g_A\hspace{2ex} \longrightarrow\hspace{2ex} 
\Delta\Sigma^{\ast} < \Delta\Sigma \,.
\label{gA}
\end{equation}
\item[(ii)]
The Pauli form factor 
in medium, $F^{\ast}_2(t)$, is {\it enhanced} relative to that 
in the vacuum, $F_2(t)$,
while the Dirac form factor remains almost the same ($F^{\ast}_1(t) \simeq F_1(t)$ for $|t|<2$ GeV$^2$) 
because of the charge conservation 
($F^{\ast}_1(0)= F_1(0)$)~\cite{QMCneuA,Horikawa:2005dh}.
Recalling the model-independent connection between the elastic form factors and the corresponding 
generalized parton distributions~\cite{Goeke:2001tz,Diehl:2003ny},
\begin{eqnarray}
\int^{1}_{-1} dx\, H^{q/N}(x,\xi,t)=F_1^{q/N}(t) \,, \quad \int^{1}_{-1} dx\, H^{q^{\ast}/N}(x,\xi,t)=F_1^{q^{\ast}/N}(t) \,,
\nonumber\\
\int^{1}_{-1} dx \, E^{q/N}(x,\xi,t)=F_2^{q/N}(t)  \,, \quad \int^{1}_{-1} dx \, E^{q^{\ast}/N}(x,\xi,t)=F_2^{q^{\ast}/N}(t) \,,
\label{eq:ff_gpd}
\end{eqnarray}
the above observations imply
\begin{equation}
F^{\ast}_1(t) \simeq F_1(t)\,,\hspace{2ex} F^{\ast}_2(t) > F_2(t) \hspace{2ex} 
\longrightarrow\hspace{2ex} H^{q/N^{\ast}} \simeq H^{q/N}\,,\hspace{2ex} E^{q/N^{\ast}} > E^{q/N} \,,
\label{F12}
\end{equation}
where superscript $q$ denotes the quark
flavor, $H^{q/N^{\ast}}$ and $E^{q/N^{\ast}}$ are the quark GPDs
of the bound nucleon, and
$F_1^{q/N^{\ast}}(t)$ and $F_2^{q/N^{\ast}}(t)$ are the contributions of quark flavor $q$ to the
elastic Dirac and Pauli form factors of the bound nucleon, respectively.
The corresponding quantities without an asterisk refer to the free nucleon.
\end{itemize}
Inserting the relations of Eqs.~(\ref{gA}) and~(\ref{F12}) 
in the proton spin decomposition relation~\cite{Ji:1998pc} for
the in-medium and vacuum cases and summing over the quark flavors, we obtain
\begin{align}
J^{q^{\ast}} &=\frac{1}{2} - J^{g^{\ast}} = \Delta\Sigma^{\ast} + L^{q^{\ast}}
= \lim_{t,\xi \to 0}\frac{1}{2} \sum_q \int_{-1}^1 dx\, x\, (H^{q/N^{\ast}}(x,\xi,t) + E^{q/N^{\ast}}(x,\xi,t)) \nonumber\\
&  > \lim_{t,\xi \to 0}\frac{1}{2} \sum_q \int_{-1}^1 dx\, x\, (H^{q/N}(x,\xi,t) + E^{q/N}(x,\xi,t))
= \Delta\Sigma + L^q
= 1/2 - J^g = J^q \,,
\label{spinsum1}
\end{align}
where ($J^{q^{\ast}},L^{q^{\ast}},J^{g^{\ast}}$) [($J^q,L^q,J^g$)] are the 
(net quark helicity, net quark orbital angular momentum, gluon total angular momentum) 
contribution to the proton spin in medium 
(in vacuum). 
Equation~(\ref{spinsum1}) demonstrates that $J^{q^{\ast}} > J^q$ and 
$J^{g^{\ast}} < J^g$. In addition, using the fact that $\Delta\Sigma^{\ast} < \Delta\Sigma$, Eq.~(\ref{spinsum1}) leads to $L^{q^{\ast}} > L^q$.
Below, by an explicit calculation, we demonstrate that these relations are indeed
true and quantify the effect of the medium modifications.   


We assume that the quark GPDs of the bound nucleon are modified in proportion 
to the corresponding quark contribution to the bound nucleon elastic form factors,
\begin{eqnarray}
H^{q/N^{\ast}}(x,\xi,t)&=&\frac{F_1^{q/N^{\ast}}(t)}{F_1^{q/N}(t)} \,
H^{q/N}(x,\xi,t) \,, \nonumber\\
E^{q/N^{\ast}}(x,\xi,t)&=&\frac{F_2^{q/N^{\ast}}(t)}{F_2^{q/N}(t)} 
E^{q/N}(x,\xi,t) \,.
\label{eq:model}
\end{eqnarray}
By construction, our model for the bound nucleon GPDs preserves the fundamental 
property of polynomiality of the bound nucleon GPDs (provided that the free nucleon GPDs obey polynomiality), which is a consequence of Lorentz invariance and which
states that the $x$ integrals of $x^n H^{q/N^{\ast}}$ and $x^n E^{q/N^{\ast}}$
are polynomials in $\xi^2$ of order $n$ for even $n$ and of order $n+1$ for odd $n$.
As a particular example of polynomiality, 
our model for the bound nucleon GPDs is constrained to
reproduce the elastic form factors of the bound nucleon [see Eq.~(\ref{eq:ff_gpd})]:
\begin{eqnarray}
\sum_q e_q \int^{1}_{-1} dx\, H^{q/N^{\ast}}(x,\xi,t)=\sum_q e_q  \frac{F_1^{q/N^{\ast}}(t)}{F_1^{q/N}(t)} \int^{1}_{-1} dx\, H^{q/N}(x,\xi,t)=\sum_q
e_q F_1^{q/N^{\ast}} \equiv F_1^{N^{\ast}}(t) \,, \nonumber\\
\sum_q e_q \int^{1}_{-1} dx\, E^{q/N^{\ast}}(x,\xi,t)=\sum_q e_q  \frac{F_2^{q/N^{\ast}}(t)}{F_2^{q/N}(t)} \int^{1}_{-1} dx\, E^{q/N}(x,\xi,t)=\sum_q
e_q F_2^{q/N^{\ast}} \equiv F_2^{N^{\ast}}(t) \,,
\label{eq:bound_ff}
\end{eqnarray}
where $e_q$ is the electric charge of quark flavor $q$.
One should emphasize that it is  Eqs.~(\ref{eq:ff_gpd}) and (\ref{eq:bound_ff}) that
 motivated our model for the
bound nucleon GPDs in Eq.~(\ref{eq:model}).

The $t$ dependence of the bound nucleon GPDs comes from the $t$ dependence of
the free nucleon GPDs and from the $t$ dependence of the ratio of the quark contribution to
the bound and free nucleon form factors.
It is important to point out that our model of the bound nucleon GPD 
neglects the EMC, Fermi motion, nuclear shadowing and antishadowing effects. 
We estimated the reliability of this approximation
and found that the effect of this approximation on $J^{q^{\ast}}$ is small:
the EMC and nuclear shadowing effects are counterbalanced by the antishadowing and 
Fermi motion effects in the integral for  $J^{q^{\ast}}$.
For details, see the discussion below.

Provided that the strange quark contribution is small, as shown by recent parity
violation experiments~\cite{Armstrong:2005hs,Acha:2006my},
the $u$ and $d$ quark contributions to the elastic form factors of the proton and neutron,
$F_{1,2}^p(t)$ and $F_{1,2}^n(t)$, are 
\begin{eqnarray}
F_{1,2}^p(t)&=&\frac{2}{3}F_{1,2}^u(t)-\frac{1}{3}F_{1,2}^d(t) \,, \nonumber\\
F_{1,2}^n(t)&=&\frac{2}{3}F_{1,2}^d(t)-\frac{1}{3}F_{1,2}^u(t) \,,
\label{eq:quark_decomposition}
\end{eqnarray}
where
each flavor is accompanied by its electric charge.
In the second line, we used charge symmetry, which relates the quark contributions to the elastic form factors of the neutron to those of the proton,
$F_{1,2}^{u/n}(t)=F_{1,2}^{d/p}(t) \equiv F_{1,2}^{d}(t)$ and
$F_{1,2}^{d/n}(t)=F_{1,2}^{u/p}(t) \equiv F_{1,2}^{u}(t)$. 
Similar relations hold for the bound
proton and neutron.

Using Eq.~(\ref{eq:quark_decomposition}) for the bound and free nucleon, our model for the quark
GPDs of the bound proton reads
\begin{eqnarray}
H^{u/p^{\ast}}(x,\xi,t)& =& \frac{2\,F_1^{p^{\ast}}(t)+F_1^{n^{\ast}}(t)}{2\,F_1^{p}(t)+F_1^{n}(t)}
H^u(x,\xi,t)=r_1^p(t)\frac{1+\frac{1}{2} \frac{r_1^n(t)}{r_1^p(t)}
\frac{F_1^n(t)}{F_1^p(t)}}{1+\frac{1}{2}\frac{F_1^n(t)}{F_1^p(t)}} H^u(x,\xi,t)
 \,, \nonumber\\
H^{d/p^{\ast}}(x,\xi,t)& =& \frac{F_1^{p^{\ast}}(t)+2\,F_1^{n^{\ast}}(t)}{F_1^{p}(t)+2\,F_1^{n}(t)}
H^d(x,\xi,t)=r_1^p(t)\frac{1+2\,\frac{r_1^n(t)}{r_1^p(t)}
\frac{F_1^n(t)}{F_1^p(t)}}{1+2\,\frac{F_1^n(t)}{F_1^p(t)}} H^d(x,\xi,t)
\,, \nonumber\\
E^{u/p^{\ast}}(x,\xi,t)& =& \frac{2\,F_2^{p^{\ast}}(t)+F_2^{n^{\ast}}(t)}{2\,F_2^{p}(t)+F_2^{n}(t)}
E^u(x,\xi,t)=r_2^p(t)\frac{1+\frac{1}{2} \frac{r_2^n(t)}{r_2^p(t)}
\frac{F_2^n(t)}{F_2^p(t)}}{1+\frac{1}{2}\frac{F_2^n(t)}{F_2^p(t)}}
E^u(x,\xi,t)
 \,, \nonumber\\
E^{d/p^{\ast}}(x,\xi,t)& =& \frac{F_2^{p^{\ast}}(t)+2\,F_2^{n^{\ast}}(t)}{F_2^{p}(t)+2\,F_2^{n}(t)}
E^d(x,\xi,t)=r_2^p(t)\frac{1+2\,\frac{r_2^n(t)}{r_2^p(t)}
\frac{F_2^n(t)}{F_2^p(t)}}{1+2\,\frac{F_2^n(t)}{F_2^p(t)}} E^d(x,\xi,t)\,,
\label{eq:model2}
\end{eqnarray}
where we introduced the shorthand notation for the ratio of the bound to free proton 
and neutron elastic form factors,
\begin{eqnarray}
r_{1,2}^p & \equiv & \frac{F_{1,2}^{p^{\ast}}(t)}{F_{1,2}^{p}(t)} \,, \nonumber\\
r_{1,2}^n & \equiv & \frac{F_{1,2}^{n^{\ast}}(t)}{F_{1,2}^{n}(t)} \,.
\label{eq:r12}
\end{eqnarray}
Note that charge symmetry for the quark contributions 
to the nucleon elastic form factors and for the free nucleon 
GPDs leads to charge symmetry for the bound nucleon GPDs [see 
Eq.~(\ref{eq:model})]. Therefore, 
\begin{eqnarray}
H^{u/n^{\ast}}(x,\xi,t)=H^{d/p^{\ast}}(x,\xi,t) \,,\nonumber\\
H^{d/n^{\ast}}(x,\xi,t)=H^{u/p^{\ast}}(x,\xi,t) \,, \nonumber\\
E^{u/n^{\ast}}(x,\xi,t)=E^{d/p^{\ast}}(x,\xi,t) \,,\nonumber\\
E^{d/n^{\ast}}(x,\xi,t)=E^{u/p^{\ast}}(x,\xi,t) \,,
\label{eq:neutron}
\end{eqnarray}
where the right-hand side of Eq.~(\ref{eq:neutron}) is given by Eq.~(\ref{eq:model2}).
In addition, we assume that the strange quark GPDs 
are not modified by the nuclear medium, e.g.~$H^{s/p^{\ast}}(x,\xi,t)=H^{s/n^{\ast}}(x,\xi,t)=H^s(x,\xi,t)$.
Note also that the model used in our recent analysis of the bound nucleon GPDs in 
$^4$He~\cite{Guzey:2008fe} is slightly different from our present model given by Eq.~(\ref{eq:model2}) and leads to small violations of charge symmetry for the bound nucleon.

In the forward limit, which is relevant for the Ji spin sum rule~\cite{Ji:1996ek},  the bound proton GPDs given by Eq.~(\ref{eq:model2}) become
\begin{eqnarray}
H^{u/p^{\ast}}(x,0,0)& =& u(x)\,, \nonumber\\
H^{d/p^{\ast}}(x,0,0)& =& d(x) \,, \nonumber\\
E^{u/p^{\ast}}(x,0,0)& =& r_2^p(0)\frac{1+\frac{1}{2} \frac{r_2^n(0)}{r_2^p(0)}
\frac{k^n}{k^p}}{1+\frac{1}{2}\frac{k^n}{k^p}}
e^u(x)=\frac{2\,k^p r_2^p(0)+k^n r_2^n(0)}{2\,k^p+k^n} e^u(x) \equiv
r^u\,e^u(x)
 \,, \nonumber\\
E^{d/p^{\ast}}(x,0,0)& =& r_2^p(0)\frac{1+2\, \frac{r_2^n(0)}{r_2^p(0)}
\frac{k^n}{k^p}}{1+2\,\frac{k^n}{k^p}}
e^d(x)=\frac{k^p r_2^p(0)+2\,k^n r_2^n(0)}{k^p+2\,k^n} e^d(x) \equiv
r^d\,e^d(x) \,,
\label{eq:fl} 
\end{eqnarray}
where $u(x)$ and $d(x)$ are the $u$-quark and $d$-quark usual parton distributions, respectively;
$e^u(x)$ and $e^d(x)$ are the forward limits of the GPD 
$E^q$ for the $u$ and $d$ quark flavors, 
respectively; $k^p=1.793$ and $k^n=-1.913$ are the proton and neutron anomalous magnetic moments. For brevity, 
we introduced the factors $r^u$ and $r^d$, which, in our model, determine 
the medium modification of  the forward  limit of the GPD $E^q$,
\begin{eqnarray}
r^u&=&\frac{2\,k^p r_2^p(0)+k^n r_2^n(0)}{2\,k^p+k^n} \,, \nonumber\\
r^d&=&\frac{k^p r_2^p(0)+2\,k^n r_2^n(0)}{k^p+2\,k^n} \,.
\label{eq:r_ud}
\end{eqnarray}
The factors $r^u$ and $r^d$ are linear combinations of the factors $r_2^p(0)$ and
$r_2^n(0)$, which characterize the modifications of the Pauli form factor of 
the nucleon at the zero momentum transfer (the modifications of the 
nucleon anomalous magnetic moment). For the latter, we used
the results of the quark-meson coupling (QMC) model~\cite{Lu:1997mu,QMCgA,QMCneuA}.
In the QMC model, medium modifications depend on the nuclear density and the effect
increases as the nuclear density is increased.

\begin{figure}[ht]
\begin{center}
\epsfig{file=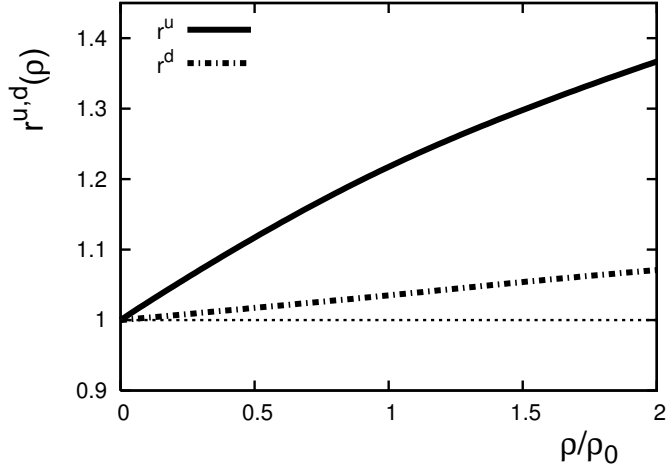,scale=1.2}
\caption{The factors $r^u$ and $r^d$, which define the medium modification of 
the forward  limit of the GPD $E^q$ in our model, see Eqs.~(\ref{eq:fl}), (\ref{eq:r_ud}) and (\ref{eq:r12}),
as a function of $\rho/\rho_0$, where  $\rho$ is the nuclear density and
$\rho_0=0.15$ fm$^{-3}$.
The medium modifications are calculated using the results of the 
QMC model~\cite{Lu:1997mu,QMCgA,QMCneuA}.}
\label{fig:r_ud}
\end{center}
\end{figure}

Figure~\ref{fig:r_ud} presents the factors $r^u$ and $r^d$
as a function of $\rho/\rho_0$, where  $\rho$ is the nuclear density and
$\rho_0=0.15$ fm$^{-3}$ is the density of the nuclear matter.
Note that the nuclear density at the center of sufficiently heavy nuclei
is close to $\rho_0$.

In addition, for the forward limit of the nucleon GPDs~(\ref{eq:fl}),
we used the following input.
The quark parton distributions (PDFs) were taken from the next-to-next-to-leading
order (NNLO) parametrization by MRST2002 at $Q^2=1$ GeV$^2$,
which corresponds to three quark flavors~\cite{Martin:2002dr}.
For the forward limit of the GPD $E^q$ denoted by $e^u(x)$ and $e^d(x)$,  we used the
model of Ref.~\cite{Guidal:2004nd}, which provides a good description of the free
proton and neutron elastic form factors,
\begin{eqnarray}
e^u(x) & =& \frac{k^u}{N_u}\,(1-x)^{\eta_u} u_v(x) \,,
\nonumber\\
e^d(x) & =& \frac{k^d}{N_d}\,(1-x)^{\eta_d} d_v(x) \,,
\label{eq:model_e}
\end{eqnarray}
where 
$k^u=2\,k^p+k^n=1.673$ and $k^d=k^p+2\,k^n=-2.033$ are the quark contributions to the 
nucleon anomalous magnetic moment and $u_v(x)$ and $d_v(x)$ are the
$u$ and $d$ valence quark distributions, respectively.
The free parameters $\eta_u=1.713$ and $\eta_d=0.566$ are determined from fits to the nucleon elastic form factors. $N_u$ and $N_d$ are the normalization factors,
$N_u = \int^1_0 dx\, (1-x)^{\eta_u} u_v(x)$ and
$N_d = \int^1_0 dx\, (1-x)^{\eta_d} d_v(x)$.
Finally, we assume that the strange quark $e^s(x)=0$. 

Note that the use of the NNLO MRST2002 parametrization for the quark 
distributions and the resolution  scale $Q^2=1$ GeV$^2$ as well as the model for
$e^u(x)$ and $e^d(x)$ should be considered as parts of a bigger model~\cite{Guidal:2004nd},
whose parameters were
adjusted to give the best description of the nucleon (proton and neutron) elastic form factors.

Having fully specified our model for the forward limit of the bound nucleon GPDs,
we can examine the influence of the medium modifications
on the spin sum rule for the bound nucleon.
The quark contribution to the bound proton spin sum rule reads
\begin{eqnarray}
2J^{q^{\ast}}&=&\sum_{q=u,d,s} \int^1_{-1} dx \, x\, \left(H^{q/p^{\ast}}(x,0,0)
+E^{q/p^{\ast}}(x,0,0)\right) \nonumber\\
&=&\sum_{q=u,d,s} \int^1_{0} dx \, x\, \left(q(x)+\bar{q}(x)+r^u e^u(x)+r^d e^d(x)
\right) \nonumber\\
&=&0.654+0.219\,r_u-0.263\,r_d \,.
\label{eq:2J}
\end{eqnarray}
Note that quark contribution to the bound neutron spin sum rule is given by the same
expression.

\begin{figure}[t]
\begin{center}
\epsfig{file=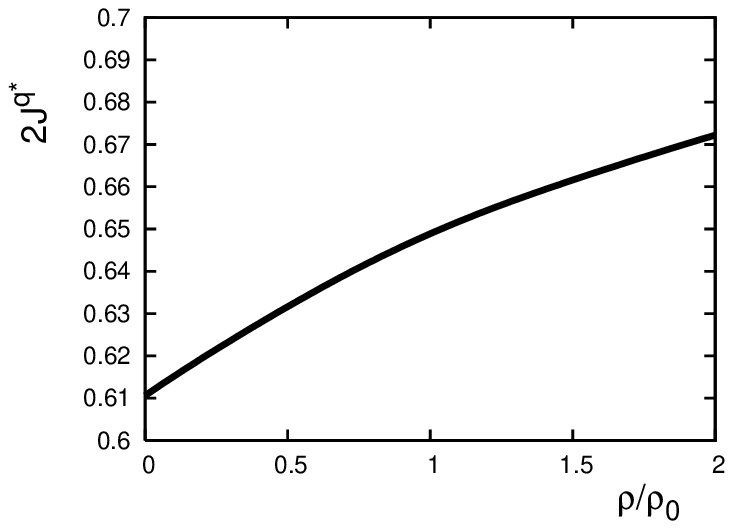,scale=1.2}
\caption{The quark contribution to the spin sum rule of the bound nucleon, $2 J^{q^{\ast}}$, 
as a function of $\rho/\rho_0$
 at $Q^2=1$ GeV$^2$, where  $\rho$ is the nuclear density and
$\rho_0=0.15$ fm$^{-3}$.
The medium modifications are calculated using the results of the QMC model.
}
\label{fig:2J}
\end{center}
\end{figure}

Figure~\ref{fig:2J} presents the quark contribution to the spin of the bound nucleon, $2 J^{q^{\ast}}$, as a function of the nuclear density at $Q^2=1$ GeV$^2$. 
The case of the free proton corresponds to $\rho/\rho_0=0$, for which
$2 J^q=0.610$.
As one can see from Fig.~\ref{fig:2J} and also from Eq.~(\ref{eq:2J}), 
the medium modifications of the bound nucleon GPD $E^{q/N^{\ast}}$ 
increase the quark contribution 
to the bound nucleon spin sum rule.

\begin{figure}[ht]
\begin{center}
\epsfig{file=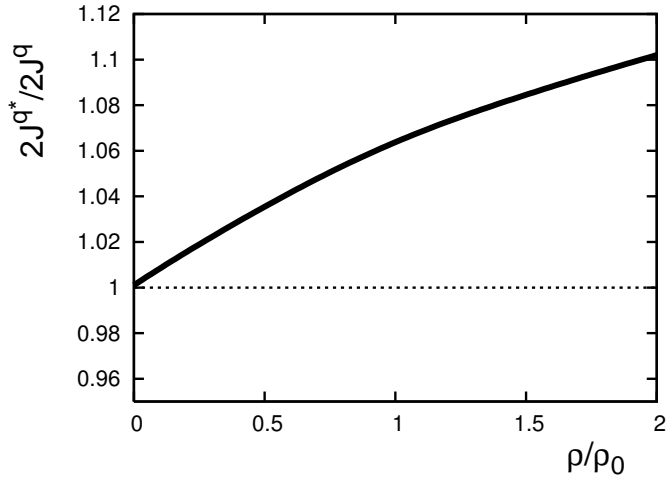,scale=1.2}
\caption{The ratio of the quark contribution 
to the bound nucleon spin sum rule to that of the free nucleon, 
$2J^{q^{\ast}}/2J^q$, as a function of $\rho/\rho_0$
 at $Q^2=1$ GeV$^2$, where  $\rho$ is the nuclear density and
$\rho_0=0.15$ fm$^{-3}$.
The medium modifications are calculated using the results of the 
QMC model.
}
\label{fig:2J_ratio}
\end{center}
\end{figure}

The effect is quite noticeable and increases with increasing 
nuclear density $\rho$. This is illustrated in Fig.~\ref{fig:2J_ratio}, where
the ratio of the quark contribution 
to the bound nucleon spin sum rule to that of the free nucleon, 
$2J^{q^{\ast}}/2J^q$, is plotted as a function of $\rho/\rho_0$.
As one can see from the figure, for instance, at $\rho=\rho_0=0.15$ fm$^{-3}$,
$2J^{q^{\ast}}/2J^q=1.070$, i.e., it is a 7\% effect.
Because the sum of the net quark and gluon contributions to the bound nucleon
spin should be one half, the gluon contribution to the bound nucleon spin sum rule,
$J^{g^{\ast}} \equiv 1/2-J^{q^{\ast}}$, is decreased in the nuclear medium.

It is important to point out that our model of the bound nucleon GPD $H^q$ has the usual
unmodified quark distribution $q(x)$ as a forward limit. This is an approximation
that neglects the Fermi motion effect and possible medium modifications of the shape of 
$q(x)$ (see the relevant discussion in Ref.~\cite{Guzey:2008fe}).
An estimate of the reliability of this approximation,
based on a parametrization of the EMC effect~\cite{Eskola:1998df},
suggests that the effect of this approximation on $2J^{q^{\ast}}$ is small:
even for a nucleus as heavy as $^{208}$Pb,
the contributions of $e^{u}$ and $e^{d}$ do not change and the contribution of
$q(x)+{\bar q}(x)$ changes (decreases) by less than 2\% (the EMC and nuclear shadowing
effects almost exactly counterbalance the Fermi motion and antishadowing effects).

The quark contribution to the spin sum rule, $J^q$, can be separated in a gauge-invariant way into the contribution of the quark helicity distributions, 
$\Delta \Sigma$, and the contribution
of the quark angular momentum, $L^q$~\cite{Ji:1996ek}. Thus, for the bound nucleon,
\begin{equation}
J^{q^{\ast}}=\Delta \Sigma^{\ast}+L^{q^{\ast}} \,,
\label{eq:decomposition}
\end{equation}
where
the quark helicity contribution to the bound nucleon spin is given by the sum 
of the first moments of the quark helicity distributions in the bound nucleon,
$\Delta q^{\ast}(x)$,
\begin{equation}
\Delta \Sigma^{\ast}=\frac{1}{2}\sum_{q=u,d,s} \int^1_0 dx (\Delta q^{\ast}(x)+\Delta {\bar q}^{\ast}(x)) \,.
\end{equation}
To estimate $\Delta \Sigma^{\ast}$, 
we assume that the contribution of the $u$ and $d$ quarks to $\Delta \Sigma^{\ast}$
is  modified (suppressed) in proportion to the medium modifications of the
axial coupling constant $g_A$ and that the contribution of the strange quark is unmodified,
\begin{equation}
\Delta \Sigma^{\ast}=\frac{g_A^{\ast}}{g_A}\, \frac{1}{2}\sum_{q=u,d} \int^1_0 dx (\Delta q(x)+\Delta {\bar q}(x))
+\frac{1}{2} \int^1_0 dx (\Delta s(x)+\Delta {\bar s}(x))
\,,
\label{eq:delta_q}
\end{equation}
where $\Delta q(x)$ are the helicity distributions of the free nucleon.
A more detailed treatment of the helicity distributions of the bound 
nucleon in the framework of the QMC model, 
which leads to the same result, $\Delta q^{\ast}(x) < \Delta q(x)$, 
can be found in Ref.~\cite{Steffens:1998rw}.
The assumption of Eq.~(\ref{eq:delta_q}) is consistent with the simultaneous
suppression of the axial coupling constant $g_A^{\ast}$ and $\Delta \Sigma^{\ast}$
because of the enhancement of the lower component of the quark Dirac spinor 
in the nuclear medium [see
Eq.~(\ref{gA}) and its qualitative discussion].
Equation~(\ref{eq:delta_q}) is also consistent with the medium modifications 
of the Bjorken sum rule~\cite{Bjorken:1966jh} and, in another language, with 
the model of the medium modifications
of the GPD $\tilde{H}$ suggested in Ref.~\cite{Guzey:2008fe}.

For the medium modifications of the axial coupling constant of the bound nucleon,
we use the results of the QMC model~\cite{Lu:1997mu,QMCgA,QMCneuA}.
For the free nucleon helicity distributions $\Delta q(x)$, we used the next-to-leading order (NLO)
GRSV2000 parametrization at $Q^2=1$ GeV$^2$~\cite{Gluck:2000dy}. 

\begin{figure}[h]
\begin{center}
\epsfig{file=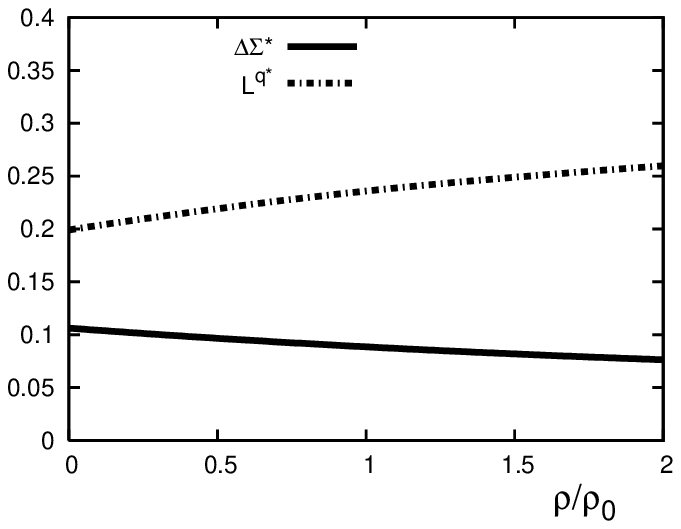,scale=1.2}
\epsfig{file=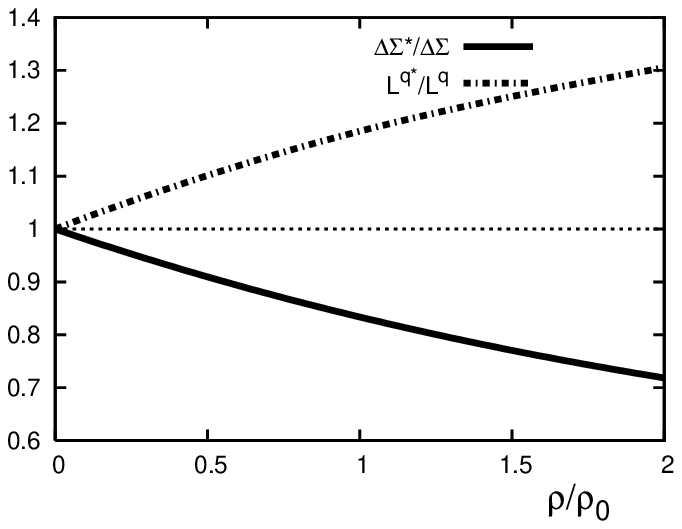,scale=1.2}
\caption{The nuclear medium modifications of the quark helicity contribution,
$\Delta \Sigma^{\ast}$, and the quark angular momentum
contribution, $L^{q^{\ast}} \equiv J^{q^{\ast}}-\Delta \Sigma^{\ast}$,
 to the bound nucleon spin  as a function of $\rho/\rho_0$ 
at $Q^2=1$ GeV$^2$, where  $\rho$ is the nuclear density and
$\rho_0=0.15$ fm$^{-3}$.
The upper panel represents the absolute values; the lower panel gives
the ratios with respect to corresponding free nucleon $\Delta \Sigma$
and $L^{q}$.
The medium modifications are calculated using the results of the 
QMC model.
}
\label{fig:Sigma}
\end{center}
\end{figure}

Figure~\ref{fig:Sigma} presents
the nuclear medium modifications of the quark helicity contribution,
$\Delta \Sigma^{\ast}$, and the quark angular momentum
contribution, $L^{q^{\ast}} \equiv J^{q^{\ast}}-\Delta \Sigma^{\ast}$,
 to the bound nucleon spin  as a function of $\rho/\rho_0$.
The upper panel represents the absolute values; the lower panel gives
the ratios with respect to the corresponding free nucleon $\Delta \Sigma$
and $L^{q}$.

As one can see from Fig.~\ref{fig:Sigma}, because of the quenching of the axial coupling
constant in the nuclear medium, $\Delta \Sigma^{\ast} < \Delta \Sigma$. As a consequence
of the relation $L^{q^{\ast}} \equiv J^{q^{\ast}}-\Delta \Sigma^{\ast}$ and the fact that
$J^{q^{\ast}} > J^{q}$, the quark angular momentum
contribution to the nucleon spin is larger for the bound nucleon compared to that for the
free nucleon, $L^{q^{\ast}} > L^q$. Both effects are large: at $\rho=\rho_0=0.15$ fm$^{-3}$,
$\Delta \Sigma^{\ast}/ \Delta \Sigma=0.83$ and $L^{q^{\ast}}/ L^q=1.20$, i.e., these
are 17\% and 20\% effects, respectively.

In summary, assuming that the bound nucleon GPDs are modified in proportion to 
the corresponding quark contributions to the bound nucleon elastic form factors,
we estimated the nuclear medium modifications
of the quark contribution to the bound nucleon spin sum rule, $J^{q^{\ast}}$,
as well the separate helicity, $\Delta \Sigma^{\ast}$, and the angular 
momentum, $L^{q^{\ast}}$,
contributions to $J^{q^{\ast}}$.
For the bound nucleon elastic form factors, we used the results of the 
quark-meson coupling model.
The resulting model of the bound nucleon GPDs
is relevant for incoherent DVCS (with nuclear breakup) with nuclear targets.
We found that the medium modifications increase  $J^{q^{\ast}}$ and $L^{q^{\ast}}$
and decrease $\Delta \Sigma^{\ast}$ compared 
to the free nucleon case. The effect is large and increases 
with increasing nuclear density $\rho$.
For instance, at $\rho=\rho_0=0.15$ fm$^{-3}$, 
$J^{q^{\ast}}$ increases by 7\%,  $L^{q^{\ast}}$ increases by 20\%, and
$\Delta \Sigma^{\ast}$ decreases by 17\%.
These effects are a general feature of relativistic mean-field quark models
and may be qualitatively explained by the enhancement 
of the lower component of the quark wave function of the bound nucleon.
 
\acknowledgments
This work was
authored by Jefferson Science Associates, LLC under U.S. DOE Contract No. DE-AC05-06OR23177.


\begin{thebibliography}{99}


\bibitem{EMC}
  J.~J.~Aubert {\it et al.}  [European Muon Collaboration],
  Phys.\ Lett.\  B {\bf 123}, 275 (1983).

\bibitem{Frankfurt:1988nt}
  L.~L.~Frankfurt and M.~I.~Strikman,
  Phys.\ Rept.\  {\bf 160}, 235 (1988).

\bibitem{Arneodo:1992wf}
  M.~Arneodo,
  Phys.\ Rept.\  {\bf 240}, 301 (1994).

\bibitem{Geesaman:1995yd}
  D.~F.~Geesaman, K.~Saito and A.~W.~Thomas,
  Ann.\ Rev.\ Nucl.\ Part.\ Sci.\  {\bf 45}, 337 (1995).

\bibitem{Piller:1999wx}
  G.~Piller and W.~Weise,
  Phys.\ Rept.\  {\bf 330}, 1 (2000).

\bibitem{Guzey:1999rq}
  V.~Guzey and M.~Strikman,
  Phys.\ Rev.\  C {\bf 61}, 014002 (1999).

\bibitem{Bissey:2001cw}
  F.~R.~P.~Bissey, V.~A.~Guzey, M.~Strikman and A.~W.~Thomas,
  Phys.\ Rev.\  C {\bf 65}, 064317 (2002).

\bibitem{Cloet:2006bq}
  I.~C.~Cloet, W.~Bentz and A.~W.~Thomas,
  Phys.\ Lett.\  B {\bf 642}, 210 (2006).

\bibitem{Strauch:2002wu}
  S.~Strauch {\it et al.}  [Jefferson Lab E93-049 Collaboration],
  Phys.\ Rev.\ Lett.\  {\bf 91}, 052301 (2003).

\bibitem{Malace:2008gf}
  S.~Malace, M.~Paolone and S.~Strauch  [Jefferson Lab Hall A Collaboration],
  AIP Conf.\ Proc.\  {\bf 1056}, 141 (2008)
  [arXiv:0807.2252 [nucl-ex]].

\bibitem{Lu:1997mu}
  D.~H.~Lu, A.~W.~Thomas, K.~Tsushima, A.~G.~Williams and K.~Saito,
  Phys.\ Lett.\  B {\bf 417}, 217 (1998);
  Phys.\ Rev.\  C {\bf 60}, 068201 (1999).

\bibitem{Schiavilla:2004xa}
  R.~Schiavilla, O.~Benhar, A.~Kievsky, L.~E.~Marcucci and M.~Viviani,
  Phys.\ Rev.\ Lett.\  {\bf 94}, 072303 (2005).

\bibitem{Saito:2005rv}
  K.~Saito, K.~Tsushima and A.~W.~Thomas,
  Prog.\ Part.\ Nucl.\ Phys.\  {\bf 58}, 1 (2007).

\bibitem{Ji:1998pc}
  X.~D.~Ji,
  J.\ Phys.\ G {\bf 24}, 1181 (1998).

\bibitem{Goeke:2001tz}
  K.~Goeke, M.~V.~Polyakov and M.~Vanderhaeghen,
  Prog.\ Part.\ Nucl.\ Phys.\  {\bf 47}, 401 (2001).

\bibitem{Diehl:2003ny}
  M.~Diehl,
  Phys.\ Rept.\  {\bf 388}, 41 (2003).

\bibitem{Belitsky:2005qn}
  A.~V.~Belitsky and A.~V.~Radyushkin,
  Phys.\ Rept.\  {\bf 418}, 1 (2005).

\bibitem{Liuti:2005gi}
  S.~Liuti and S.~K.~Taneja,
  Phys.\ Rev.\  C {\bf 72}, 032201(R) (2005);
  Phys.\ Rev.\  C {\bf 72}, 034902 (2005).

\bibitem{Liuti:2006dx}
  S.~Liuti,
  arXiv:hep-ph/0601125.

\bibitem{Guzey:2008fe}
  V.~Guzey, A.~W.~Thomas and K.~Tsushima,
Phys.\ Lett.\ B {\bf 673}, 9 (2009).

\bibitem{Ji:1996ek}
  X.~D.~Ji,
  Phys.\ Rev.\ Lett.\  {\bf 78}, 610 (1997).

\bibitem{Buck:1975ae}
  B.~Buck and S.~m.~Perez,
  Phys.\ Rev.\ Lett.\  {\bf 50}, 1975 (1983).

\bibitem{Goodman:1985xi}
  C.~D.~Goodman {\it et al.},
  Phys.\ Rev.\ Lett.\  {\bf 54}, 877 (1985).

\bibitem{Wilkinson:1973zz}
  D.~H.~Wilkinson,
  Phys.\ Rev.\  C {\bf 7}, 930 (1973).

\bibitem{Chou:1993zz}
  W.~T.~Chou, E.~K.~Warburton and B.~A.~Brown,
  Phys.\ Rev.\  C {\bf 47}, 163 (1993).


\bibitem{QMCgA}
  D.~H.~Lu, A.~W.~Thomas and K.~Tsushima,
  arXiv:nucl-th/0112001.

\bibitem{QMCneuA}
  K.~Tsushima, H.~C.~Kim and K.~Saito,
  Phys.\ Rev.\  C {\bf 70}, 038501 (2004).

\bibitem{Vaintraub:2009mm}
  S.~Vaintraub, N.~Barnea and D.~Gazit,
  arXiv:0903.1048 [nucl-th].


\bibitem{Horikawa:2005dh}
  T.~Horikawa and W.~Bentz,
  Nucl.\ Phys.\  A {\bf 762}, 102 (2005).

\bibitem{Armstrong:2005hs}
  D.~S.~Armstrong {\it et al.}  [G0 Collaboration],
  Phys.\ Rev.\ Lett.\  {\bf 95}, 092001 (2005).

\bibitem{Acha:2006my}
  A.~Acha {\it et al.}  [HAPPEX collaboration],
  Phys.\ Rev.\ Lett.\  {\bf 98}, 032301 (2007).


\bibitem{Martin:2002dr}
  A.~D.~Martin, R.~G.~Roberts, W.~J.~Stirling and R.~S.~Thorne,
  Phys.\ Lett.\  B {\bf 531}, 216 (2002).

\bibitem{Guidal:2004nd}
  M.~Guidal, M.~V.~Polyakov, A.~V.~Radyushkin and M.~Vanderhaeghen,
  Phys.\ Rev.\  D {\bf 72}, 054013 (2005).

\bibitem{Eskola:1998df}
  K.~J.~Eskola, V.~J.~Kolhinen and C.~A.~Salgado,
  Eur.\ Phys.\ J.\  C {\bf 9}, 61 (1999).


\bibitem{Steffens:1998rw}
  F.~M.~Steffens, K.~Tsushima, A.~W.~Thomas and K.~Saito,
  Phys.\ Lett.\  B {\bf 447}, 233 (1999).

\bibitem{Bjorken:1966jh}
  J.~D.~Bjorken,
  Phys.\ Rev.\  {\bf 148}, 1467 (1966).

\bibitem{Gluck:2000dy}
  M.~Gluck, E.~Reya, M.~Stratmann and W.~Vogelsang,
  Phys.\ Rev.\  D {\bf 63}, 094005 (2001).







\end{thebibliography}
\end{document}